\documentclass[10pt,conference]{IEEEtran}
\usepackage[utf8]{inputenc}
\usepackage[T1]{fontenc}
\usepackage{amsmath,amssymb}
\usepackage{textcomp}
\usepackage{graphicx}
\usepackage{booktabs}
\usepackage{cite}
\usepackage{url}
\usepackage{xcolor}
\usepackage{multirow}
\usepackage[hidelinks]{hyperref}
\usepackage{xurl}
\usepackage{tikz}
\usetikzlibrary{trees,positioning}

\begin{document}

\title{Hierarchical Decoding for Quantum Error Correction}

\author{\parbox{0.96\textwidth}{\centering
{\fontsize{11}{14}\selectfont
\mbox{Xulong Tang\textsuperscript{1}},
\mbox{Narges Alavisamani\textsuperscript{2}},
\mbox{Joan Camps\textsuperscript{3}},
\mbox{Matthew Cha\textsuperscript{4}},
\mbox{Elise Jennings\textsuperscript{3}},
\mbox{Jack O'Connor\textsuperscript{3}},
\mbox{Humberto Munoz Bauza\textsuperscript{2}},
\mbox{Josh Izaac\textsuperscript{5}},
\mbox{Namit Anand\textsuperscript{2}},
\mbox{Ron Bewtra\textsuperscript{6}},
\mbox{Jeremiah Coleman\textsuperscript{6}},
\mbox{Johannes Heinsoo\textsuperscript{6}},
\mbox{Michael Ferguson\textsuperscript{2}},
\mbox{Ben Lau\textsuperscript{5}},
\mbox{P. Aaron Lott\textsuperscript{2}},
\mbox{Leonhard Neuhaus\textsuperscript{5}},
\mbox{Lee J. O'Riordan\textsuperscript{5}},
\mbox{Muhammad Osama\textsuperscript{1}},
\mbox{Masoud Mohseni\textsuperscript{2}},
\mbox{Blake R. Johnson\textsuperscript{7}}, and
\mbox{Yasuko Eckert\textsuperscript{1}}\par}
\vspace{0.6em}
{\fontsize{10}{12}\selectfont
\textsuperscript{1}Advanced Micro Devices, Inc., Research and Advanced Development, USA\\
\textsuperscript{2}HPE Quantum, Emergent Machine Intelligence, HPE Labs, USA\\
\textsuperscript{3}Riverlane, Cambridge, CB2 3BZ, UK\\
\textsuperscript{4}Riverlane Research Inc., Cambridge, MA, 02142, USA\\
\textsuperscript{5}Xanadu, Toronto, ON, M5G 2C8, Canada\\
\textsuperscript{6}IQM Quantum Computers, Riverdale, MD, USA\\
\textsuperscript{7}IBM Research, Yorktown Heights, NY, USA\par}
}
}

\maketitle

\begin{abstract}
Fault-tolerant quantum computing requires classical decoders that can keep up with syndrome data streams and supply correction information and logical measurement outcomes to the quantum control system. Meeting these requirements at scale requires high throughput while controlling the end-to-end latency from readout to usable correction. In this paper, we organize decoding algorithms, execution strategies, and code constructions using common classifications and examine their performance and trade-offs under the reported evaluation conditions. We distinguish decoder-level hierarchy, where multiple decoding stages or resources cooperate to decode a single code, from code-level hierarchy, where the code construction itself introduces multiple decoding levels. We first review single-code hierarchical decoders and their cascaded predecoding and windowed, streaming, and distributed implementations. We then study hierarchical and concatenated codes, focusing on how decoding is coordinated across code levels through hard decisions, soft reliability information, or syndrome information. We also examine how qubit modalities shape decoding workloads and timing requirements, and review how different decoding platforms meet these requirements, including central processing units (CPUs), graphics processing units (GPUs), field-programmable gate arrays (FPGAs), application-specific integrated circuits (ASICs), and cryogenic platforms. By assessing design trade-offs, this paper clarifies connections across the literature and provides a shared foundation for QEC design exploration. It also identifies gaps in performance reporting and system integration to guide future research toward practical quantum error correction.
\end{abstract}


\section{Introduction}
\label{sec:intro}
Solving utility-scale problems with quantum computers requires suppressing logical errors throughout the computation~\cite{mohseni2026buildquantumsupercomputerscaling}. The leading framework for this is fault-tolerant quantum computing (FTQC), which encodes logical information into physical qubits and repeatedly measures parity checks. This produces a stream of syndrome data that a quantum error correction (QEC) decoder interprets to infer errors and determine appropriate corrections~\cite{Fowler2012}. To approach a low logical error rate, the decoder must process this syndrome stream in a timely manner and provide correction information when required by the quantum control system. This motivates real-time decoding as an integral part of FTQC.

Real-time decoding has two related requirements: sustained throughput and end-to-end latency. Sustained decoding throughput is the amount of syndrome data a decoding system can process per unit time during continuous operation. It must keep pace with syndrome generation to prevent a growing syndrome backlog. 
End-to-end readout-to-correction latency is the elapsed time from the start of the relevant qubit readout until the resulting correction information becomes usable by the quantum control system. It includes readout, data transport, queuing, decoding, and control processing, with decoder runtime contributing part of the total. This latency matters when a subsequent logical operation depends on the decoded result: a delayed result can stall that operation and increase storage-error exposure~\cite{Khalid2025,RTStack2026}. 
It should be noted that, in general, the aforementioned latencies vary across different syndrome rounds. For example, decoding latency varies when different syndrome patterns require different amounts of processing (e.g., the number of iterations in belief propagation decoding). Also, the data transport latency depends on the available communication bandwidth. When buffering and the logical-operation schedule allow temporary delays, the end-to-end latency target need not be a strict deadline for every set of syndrome data (e.g., every window in windowed decoding). Occasional overruns can be tolerated when buffering accommodates temporary delays and sufficient processing capacity allows the decoder to clear the backlog, although delays at dependent logical operations may extend the computation~\cite{Ye2026MegaQuOp}.

Alongside the above timing requirements, decoding accuracy, communication bandwidth, and hardware resource and energy costs also guide decoder design. First, accuracy matters because misidentified errors may introduce logical faults. Second, the system must provide enough communication bandwidth for large syndrome streams to flow between the quantum control system and decoder hardware. Insufficient bandwidth will cause congestion delays that increase the end-to-end readout-to-correction latency. Third, hardware resource and energy costs matter because decoding must support many logical qubits while remaining within the platform's area, power, and cooling budgets. Together, these requirements create trade-offs in decoder design. 

\begin{figure*}[!t]
\centering
\begin{tikzpicture}[
 font=\footnotesize,
 box/.style={draw,rounded corners,align=left,text width=7.0cm,inner sep=6pt},
 banner/.style={draw,rounded corners,fill=black!7,align=center,text width=15.0cm,inner sep=6pt}]
\node[banner] at (0,0) {\textbf{Decoder design space}: choose a code and decoding method, then evaluate accuracy, throughput, readout-to-correction latency, bandwidth, and cost.};
\node[box,anchor=north west] at (-7.6,-0.8) {\textbf{Single-code decoding (Section~\ref{sec:engines})}\\
Complete single-decoder algorithms\\
$\quad$Matching, clustering, BP, learned methods\\[3pt]
\textbf{Extensions and execution choices}\\
Predecoding: preprocessing / optional fallback\\
Windowing and streaming: bounded temporal state\\
Distributed execution: flat workers or merge layers\\[3pt]
The code need not change; a separate global tier is optional.};
\node[box,anchor=north west] at (0.2,-0.8) {\textbf{Hierarchical codes (Section~\ref{sec:codes})}\\
Concatenated inner and outer encoding layers\\
$\quad$Block, surface, qLDPC, or bosonic codes\\[3pt]
\textbf{Decoding across code layers}\\
Hard decisions or erasure flags\\
Soft logical-error information\\
Separately extracted outer-code syndromes\\[3pt]
Code structure and decoding interface are designed together.};
\node[banner] at (0,-4.6) {\textbf{Implementation (Section~\ref{sec:systems})}: CPU, GPU, FPGA, ASIC, and cryogenic options for either branch; readout, syndrome formation, transport, queues, decoding, and control contribute to response time.};
\end{tikzpicture}
\caption{Paper organization: single-code decoders and their execution strategies, concatenated codes and their decoders, and hardware implementations of both.}
\label{fig:taxonomy}
\end{figure*}
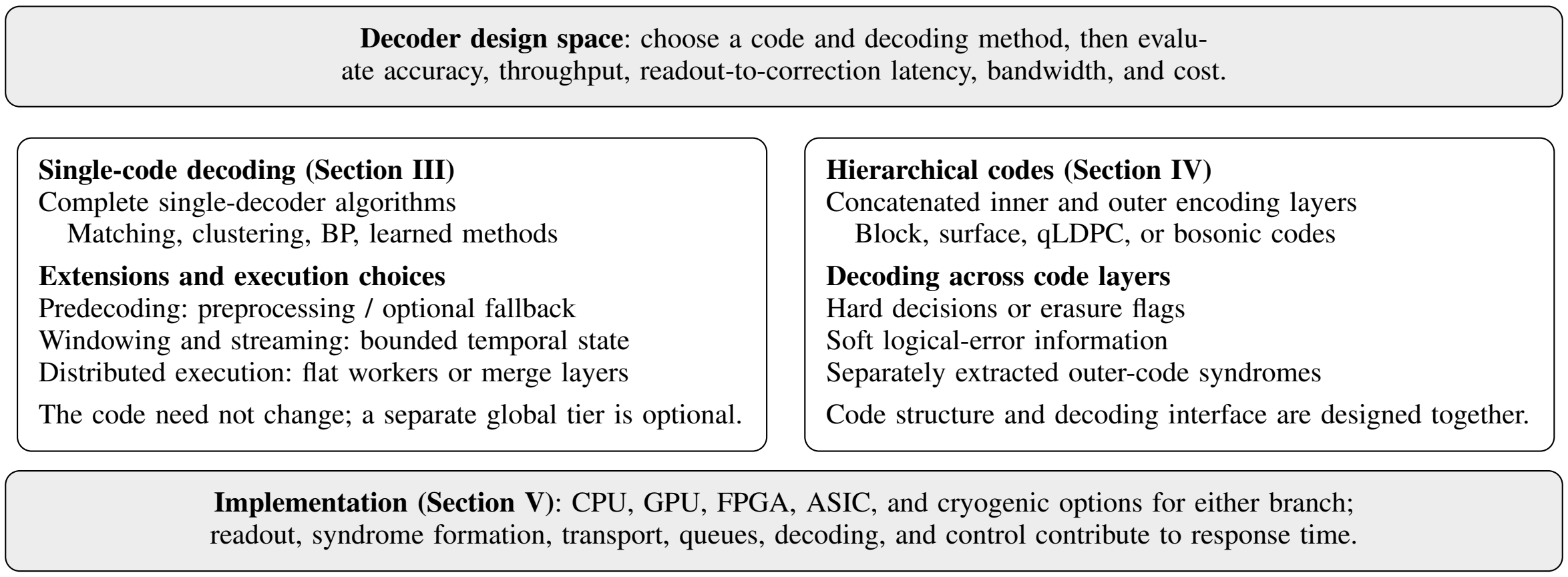

The rich decoding literature has explored these requirements and their trade-offs through a wide range of algorithms, code constructions, and implementations. One line of research studies decoders for a single code and explores hierarchical or multi-tier implementations and deployment strategies, including cascaded predecoding~\cite{Delfosse2020,Chamberland2022,Chamberland2026Pre} and windowed, streaming, and distributed execution~\cite{Huang2023,Skoric2022,LATTE2025}. These approaches organize decoding work across stages, syndrome regions, or hardware components. Another line of research explores hierarchical code families, including concatenated codes, and their corresponding decoders, where decoding follows the encoding layers and exchanges information between inner and outer decoders~\cite{Poulin2006,Meister2024}. Figure~\ref{fig:taxonomy} shows these two code settings and the hardware platforms used to implement their decoders.

Many of these decoding approaches are evaluated under different qubit modalities, code sizes, noise models, and classical hardware platforms. Definitions and measurement boundaries for throughput and latency also vary across studies, limiting direct comparison of reported results. In this paper, we summarize the reported performance and discuss the trade-offs identified in prior studies, retaining their evaluation conditions and distinguishing decoder runtime from end-to-end latency. We also examine how qubit modalities shape decoding requirements and inform the choice of classical hardware.

Specifically, this paper has three major parts. First, we review single-code decoders (e.g., Relay-BP and hardware clustering~\cite{Muller2025,Barber2023,Liyanage2024}) and explore their multi-stage and distributed implementations, such as cascaded predecoding, windowing, streaming, and spatial partitioning~\cite{Delfosse2020,Chamberland2026Pre,Huang2023,Skoric2022}. We summarize their reported benefits and limitations in decoding accuracy, throughput, communication, and resource use, including preprocessing and coordination overhead. 
Second, we examine hierarchical code constructions, including concatenated codes, and their corresponding decoders~\cite{Poulin2006,Meister2024}. We discuss how decoding follows the encoding layers and uses hard decisions, erasures, soft reliability information, or outer-code syndromes, and how information exchange between layers affects the reliability and cost of the complete construction. 
Finally, we review hardware implementations on CPUs, GPUs, FPGAs, ASICs, and cryogenic platforms. We examine how qubit modalities shape syndrome-generation rates, readout overhead, and the timing of dependent logical operations, and how these requirements inform hardware choices. Across these implementations, we summarize reported decoding accuracy, throughput, end-to-end latency, communication bandwidth, resource use, and energy consumption, accounting for the code parameters, noise models, and evaluation conditions of each study.

The remainder of the paper is organized as follows. Section~\ref{sec:background} introduces decoder requirements and performance metrics. Section~\ref{sec:engines} reviews single-code algorithms and their cascaded, windowed, streaming, and distributed implementations. Section~\ref{sec:codes} examines hierarchical code constructions and their corresponding decoders. Section~\ref{sec:systems} reviews hardware implementations and their reported performance, distinguishing decoder runtime from end-to-end readout-to-correction latency. Section~\ref{sec:challenges} discusses challenges and open problems.

\section{Decoder Requirements}
\label{sec:background}
\label{sec:bg-constraints}

This section defines five requirements used to discuss reported decoding performance: throughput, end-to-end latency, decoding accuracy, communication bandwidth, and hardware resource and energy costs. We summarize the metrics reported in each study while retaining its evaluation and measurement conditions.

\subsection{Decoder requirements}

\emph{(C1) Decoding throughput}: sustained decoding throughput is the amount of syndrome data processed per unit time during continuous operation, reported in syndrome rounds or detection events per second for a specified code size and noise model. The decoder must keep pace with syndrome generation, with additional processing capacity to clear temporary backlogs. Higher physical error rates generally produce more detection events, which can increase decoding work and reduce throughput, depending on the decoding algorithm. Lower physical error rates can ease this processing demand without changing the rate at which syndrome measurements are generated. 
For example, in windowed decoding, where consecutive (typically overlapping) windows of syndrome rounds are decoded separately, a sequential decoder must process each window, on average, before enough new syndrome rounds arrive to advance to the next window~\cite{Huang2023,Skoric2022, tan2023scalable}.

\emph{(C2) End-to-end readout-to-correction latency}: let $L_{\rm e2e}$ denote the elapsed time from the start of the relevant qubit readout until the resulting correction information becomes usable by the quantum control system.
Reported latency depends on where timing starts and ends, so these measurement boundaries must be specified when comparing results from different studies.
For a serial path, a useful accounting is
\begin{equation}
\begin{split}
 L_{\rm e2e}={}&L_{\rm readout}+L_{\rm form}+L_{\rm transport}\\
              &+L_{\rm wait}+L_{\rm decode}+L_{\rm apply},
\end{split}
\label{eq:e2e-latency}
\end{equation}
where $L_{\rm readout}$ is the time to collect and interpret qubit readout signals, $L_{\rm form}$ is the time to form syndrome data, $L_{\rm transport}$ is the time to transfer syndrome data to the decoder and return correction information, $L_{\rm wait}$ is the time that syndrome data or decoded results spend waiting for available processing or communication resources, $L_{\rm decode}$ is the decoding time, and $L_{\rm apply}$ is the time for the control system to process the decoded result and make the correction information usable.
It is worth mentioning that some stages can overlap. For example, in sliding-window decoding, collecting new syndrome rounds while decoding an earlier window primarily improves throughput. Overlapping stages within the same window, such as syndrome formation and data transport, can also reduce $L_{\rm e2e}$ for that window.

In sliding-window decoding, individual windows need not all meet the same fixed deadline, provided that temporary delays can be buffered and the decoder maintains sufficient throughput to clear the backlog. During Pauli-frame tracking, correction information can be incorporated later, as long as no intervening operation requires it. When a logical operation depends on a decoded result, a delay may stall that operation and increase the time qubits remain exposed to noise~\cite{Skoric2022,Khalid2025,RTStack2026}. A late result therefore does not necessarily cause the computation to fail, although waiting for it can increase runtime and logical error probability. Variability can arise in decoding ($L_{\rm decode}$), waiting for available resources ($L_{\rm wait}$), data transfer ($L_{\rm transport}$), and control processing ($L_{\rm apply}$). $L_{\rm e2e}$ should therefore be characterized by its distribution, including the mean and how often long delays occur and how long they last.

\emph{(C3) Decoding accuracy}: decoding accuracy is evaluated by the logical error rate achieved for a given code and noise model. The decoder must keep this error rate below the target required by the application. Changes intended to improve throughput, reduce $L_{\rm e2e}$, or lower hardware cost can also affect accuracy, so these trade-offs must be evaluated together. Comparisons should specify the code parameters, physical error rate, noise model, and number of error-correction rounds.

\emph{(C4) Transmission bandwidth}: communication links must carry syndrome data from the quantum control system to the decoder fast enough to keep pace with its generation. When decoding is split across multiple components, their links must also support the data exchanged between them, such as residual syndrome data sent by a local predecoder to a downstream decoder. Bandwidth affects the data-transfer time, $L_{\rm transport}$. If data arrives faster than a link can transmit it, data accumulates and increases $L_{\rm wait}$, adding to $L_{\rm e2e}$ even when decoding itself is fast. Local preprocessing can reduce the amount of data transmitted. However, some designs must still send the full syndrome when the local stage cannot determine a correction, so the available bandwidth and buffers must accommodate these cases~\cite{Delfosse2020,Smith2022,Pinball2025,LATTE2025}.

\emph{(C5) Hardware cost}: compute, memory, circuit area, power, and energy must remain practical at the target code size and logical-qubit count~\cite{Barber2023,Kadomoto2025,Pinball2025,Wang2024Cryo}.
The available compute and memory must support continuous decoding, including storage of syndrome data and intermediate results. As the number of logical qubits grows, replicating decoder hardware or sharing processing resources must remain within the system's area, power, and cooling budgets. Additional parallel hardware can increase throughput, but also requires more resources and power. Energy use depends on both power consumption and processing time, so comparisons should specify the energy per decoding task, such as a syndrome round or window, together with the code size and operating conditions. System-level estimates should also include the host processors, communication links, and cooling needed to operate the decoder.

\subsection{Qubit modality considerations}

The five requirements above apply across qubit modalities, but syndrome-generation rates, noise characteristics, readout times, and the timing of dependent logical operations differ between physical platforms.
Table~\ref{tab:modalities} summarizes examples of these differences.

\begin{table*}[!t]
\centering
\caption{Timing requirements and hardware considerations for representative qubit platforms. Values are specific to the cited studies; n/r means not reported.}
\label{tab:modalities}
\scriptsize
\renewcommand{\arraystretch}{1.4}
\begin{tabular}{@{}p{1.9cm}p{2.2cm}p{2.6cm}p{4.2cm}p{4.2cm}@{}}
\toprule
Modality / example & Syndrome cycle & Reaction target & Representative evidence & Placement implication \\
\midrule
Superconducting / Willow~\cite{GoogleWillow2025} & $1.1\,\mu$s & Protocol-dependent; not one cycle & Live stream; $63\pm17\,\mu$s decoder-only terminal latency & High throughput needed; control, communication and decoding must be budgeted separately \\
Trapped-ion / Helios~\cite{NVIDIAHelios2025,NVIDIAHeliosNews2025} & n/r in cited report & $2$ ms, vendor-stated & GH200 BP+OSD; $67\,\mu$s median decode; vendor-reported feed-forward & Networked GPU is feasible for this evaluated workload \\
Neutral-atom transversal architectures~\cite{Cain2025Correlated,AtomArray2025} & Approximately $1$ ms in architectural estimates & $1$ ms, assumed in resource analysis & Numerical correlated-decoding study and resource estimates using one syndrome-extraction round after each transversal gate, not a measured QPU timing loop & Inter-block correlations, graph volume, and data movement shape parallel execution \\
Silicon spin / three-qubit phase code~\cite{Takeda2022Spin} & No repeated measured-syndrome cycle & No classical decoder-loop budget & Coherent iToffoli correction; no measured classical decoding path & Readout and reset must be characterized before assigning a classical decoder tier \\
\bottomrule
\end{tabular}
\end{table*}
Microsecond superconducting rounds require high sustained throughput, but do not require every decoding window or feed-forward operation to finish within one round.
Trapped-ion Helios demonstration instead used a two-millisecond decoding requirement~\cite{NVIDIAHeliosNews2025}.
For transversal architectures, including those motivated by neutral-atom systems, correlations between interacting blocks and the volume of syndrome data also affect decoding requirements~\cite{Cain2025Correlated}.
These examples describe particular processors and protocols, rather than fixed timing requirements for an entire qubit modality. Their implications for decoding hardware are discussed in Section~\ref{sec:systems}.

\section{Single-Code Decoders and Their Extensions}
\label{sec:engines}

This section reviews decoding for a fixed code: the encoding is held constant, and the design choice is how the decoding work should be organized.
We discuss minimum-weight perfect matching (MWPM), clustering, belief propagation (BP), and learned decoders, and what the code, the available syndrome information, and the noise model imply about their applicability. Quantum low-density parity-check (qLDPC) codes have sparse parity checks, which provide the graph on which message-passing decoders operate.
We then examine predecoding cascades, which use an initial stage to reduce the work passed to a downstream decoder, and windowed, streaming, and distributed execution, which organize that work across time and processing resources.
These are extensions of a complete decoder rather than competing categories: a baseline algorithm can run on its own or as a stage within either.
``Single-code'' here marks the absence of concatenated encoding layers (Section~\ref{sec:codes}), not one logical qubit or one decoding process.
Throughout, we relate these choices to the requirements of Section~\ref{sec:bg-constraints}, and Table~\ref{tab:engines} summarizes representative complete decoders while preserving the timing definitions of each source.

\begin{table*}[t]
\centering
\caption{Representative single-code decoders and their reported accuracy, timing, and implementation evidence. Timing boundaries vary; per-round values may be amortized and exclude parts of $L_{\rm e2e}$. Here $d$ is code distance, $p$ is physical error probability under the stated noise model, and $[[n,k,d]]$ denotes $n$ physical qubits encoding $k$ logical qubits with distance $d$.}
\label{tab:engines}
\renewcommand{\arraystretch}{1.3}
\scriptsize
\begin{tabular}{@{}p{2.1cm}p{1.1cm}p{2.5cm}p{4.3cm}p{4.9cm}@{}}
\toprule
Decoder & Code & Implementation / evidence & Timing or complexity (not full C2) & Workload / accuracy scope \\
\midrule
Sparse Blossom~\cite{Higgott2023} & surface & CPU software benchmark & $<1\,\mu$s per round, single core & $d=17$, circuit depolarizing $p=10^{-3}$; exact graph MWPM \\
Union-Find~\cite{Delfosse2017} & surface & algorithm analysis & almost-linear $O(n\alpha(n))$ & complexity, not a measured reaction time \\
Micro Blossom~\cite{Wu2025} & surface & CPU/FPGA prototype & $0.8\,\mu$s mean, syndrome-ready to correction & $d=13$, circuit noise $p=10^{-3}$; includes accelerator I/O \\
Collision Clustering (FPGA)~\cite{Barber2023} & surface & measured FPGA & $810$ ns/round, mean amortized over $d$ rounds & $d=21$, 881 physical qubits; source circuit model $p=10^{-3}$ \\
Collision Clustering (ASIC)~\cite{Barber2023} & surface & post-layout estimate & $240$ ns/round, amortized & $d=23$, 1057 qubits, circuit $p=10^{-3}$; projected ${\sim}8$ mW \\
Helios~\cite{Liyanage2024} & surface & measured FPGA & $11.5$ ns/round, mean amortized & $d=21$, phenomenological $p=10^{-3}$; not the circuit-noise result \\
LILLIPUT~\cite{Das2021} & small surface & FPGA lookup design & bounded table access & small-code implementation; full lookup storage grows exponentially \\
AFS~\cite{Das2022AFS} & surface & architectural simulation & simulated execution; no live-loop bound & shares union-find resources across logical qubits \\
Astrea~\cite{Vittal2023Astrea} & small surface & architectural simulation & simulated execution; no live-loop bound & parallel matching search; reported speed is not measured silicon \\
Local Clustering~\cite{LocalClustering2024} & surface & measured FPGA, simulated noise & per-round hardware evaluation & adaptive leakage handling; accuracy depends on leakage model \\
BP-OSD~\cite{Panteleev2019} & qLDPC & software / algorithm & Gaussian elimination can cost $O(n^3)$ & reference configuration must specify BP and OSD settings \\
Soft Syndrome~\cite{Raveendran2022Soft} & qLDPC & numerical decoding study & no timing comparison made here & incorporates measurement reliability into BP \\
Ambiguity Clustering~\cite{Wolanski2024} & qLDPC & Apple M2, one CPU core & $135\,\mu$s per syndrome round & gross code, circuit depolarizing $p=0.003$; comparable BP-OSD accuracy \\
LSD~\cite{Hillmann2024} & qLDPC & algorithm / software & parallel work depends on cluster size & localized on-the-fly elimination; depth is not wall-clock runtime \\
Relay-BP~\cite{Muller2025} & qLDPC & software benchmark & runtime--accuracy trade-off across BP legs & circuit-noise bicycle-code evaluations; no matrix inversion \\
FPGA Relay-BP~\cite{Maurer2025} & qLDPC & FPGA verification, synthetic syndromes & $480$ ns mean estimate/window; $40$ ns/round amortized & twelve-cycle gross window, circuit $p=10^{-3}$; $24$ ns/iteration \\
FPGA-tailored study~\cite{Maurya2025} & qLDPC & hardware-oriented evaluation & fixed cycle budgets & compares BP, OSD and generalized union-find configurations \\
Beam search~\cite{Delfosse2025} & qLDPC & Apple M3 software & $26.2\times$ lower P99.9 runtime than BP-OSD & gross, circuit $p=10^{-3}$, width 8; matched reported logical accuracy \\
BP-SF~\cite{Wang2025BPSF} & qLDPC & CPU and GPU software & mean runtime varies with $p$ and parallelism & circuit-noise gross-code comparison; eliminates Gaussian elimination \\
GPU BP~\cite{Ferraz2025} & qLDPC & RTX 4090 benchmark & $43.7\,\mu$s average including host/device transfers & $[[784,24,24]]$, fixed ten iterations; not an accuracy-matched timing guarantee \\
AlphaQubit~\cite{Bausch2024} & surface & QPU data and simulation & accuracy study; no full C2 bound & experimental accuracy and long simulated sequences are separate evidence \\
Mamba~\cite{Mamba2025} & surface & data and software simulation & state-space block scales as $O(d^2)$ & Sycamore comparison; latency-induced-noise study is simulated \\
Sum-of-squares~\cite{Basak2026} & surface / color & optimization simulations & cost depends on relaxation level & approximation to the most likely individual error consistent with the syndrome \\
\bottomrule
\end{tabular}
\end{table*}

\subsection{Single-Decoder Algorithms}
\label{sec:baseline}
\subsubsection{Surface-Code Matching and Clustering Decoders}
\label{sec:eng-surface}
Surface-code decoders employ several implementation strategies, including optimized matching, hardware-oriented clustering, and specialized small-code decoding. MWPM pairs detection events using edge weights derived from the error model. Sparse Blossom implements graph MWPM in less than one microsecond per syndrome round on one CPU core for a distance-seventeen surface code under circuit depolarizing noise at $p=10^{-3}$~\cite{Higgott2023}. Fusion Blossom is a partition-and-fuse execution method for matching, discussed with windowing in Section~\ref{sec:window}~\cite{Wu2023}.

The almost-linear-time union-find algorithm provides a foundation for many hardware-oriented clustering decoders~\cite{Delfosse2017}. Micro Blossom instead accelerates exact matching by dividing its primal and dual operations between software and an FPGA~\cite{Wu2025}. Clustering decoders, including Collision Clustering, Helios, and the local clustering decoder, map local graph operations onto parallel hardware to achieve sub-microsecond per-round $L_{\rm decode}$ under their evaluated configurations~\cite{Barber2023,Liyanage2024,LocalClustering2024}. Other designs reduce hardware cost through pipelining or time-multiplexing~\cite{Das2022AFS,Heer2023}.

For small surface codes, bounded syndrome spaces enable lookup-table or parallel-search approaches, such as LILLIPUT and Astrea~\cite{Das2021,Vittal2023Astrea}. A full syndrome lookup table grows exponentially with the number of syndrome bits; this is a small-code design point, rather than a scalable storage strategy for increasing distance. Tensor-network decoders can approximate logical-class probabilities with polynomial computational cost at fixed bond dimension~\cite{Bravyi2014}. Sampling from an equivalent Ising $p$-spin problem~\cite{Dennis2002} and integer-programming searches for the most likely individual error~\cite{Landahl2011} provide alternative decoding benchmarks. Together, these approaches expose trade-offs among decoding accuracy, $L_{\rm decode}$, generality, and hardware resource use.

Belief propagation (BP) iteratively exchanges error-probability estimates between variable and check nodes. BP and its variants have also been applied to surface-code decoding. For example, belief-matching and belief-find combine BP with MWPM and weighted union-find, respectively~\cite{higgott2023improved}. When BP fails to converge, its estimated error probabilities determine the downstream decoder’s edge weights, enabling these methods to exploit circuit-level noise correlations. BP is particularly useful for qLDPC codes because it operates directly on their sparse parity-check graphs, whereas conventional matching requires a graph structure that is not available for every qLDPC code. We discuss BP-based qLDPC decoders in the following subsection.

\subsubsection{qLDPC Decoders}
\label{sec:eng-qldpc}
BP provides a common starting point for qLDPC decoding. The accuracy-oriented BP-OSD baseline applies ordered statistics decoding (OSD) when BP does not converge to a valid correction~\cite{Panteleev2019,Roffe2020}. BP-OSD already has a two-stage fallback structure. We group it here as a complete decoding method; describing the same decoding procedure as a single algorithm or a cascade reflects a choice of terminology, not a difference in capability.

Gaussian elimination in OSD can dominate decoding time. Ambiguity clustering and localized statistics decoding restrict elimination to smaller, independently processed clusters. Ambiguity clustering reports comparable BP-OSD accuracy under circuit depolarizing noise at $p=0.003$, with a gross-code runtime of 135 microseconds per syndrome round on one Apple M2 core~\cite{Wolanski2024}. Localized statistics decoding performs on-the-fly elimination within each cluster, giving a parallel depth that scales with cluster size~\cite{Hillmann2024}.

Beam search replaces elimination with a bounded search over candidate corrections. It branches on unreliable BP decisions and retains a bounded number of candidates, providing a $26.2\times$ reduction in the 99.9th-percentile runtime while matching BP-OSD accuracy on the gross code at circuit-noise $p=10^{-3}$ with search width eight~\cite{Delfosse2025}.

BP syndrome-flip and Relay-BP modify the BP search itself to avoid matrix inversion. BP syndrome-flip (BP-SF) identifies unreliable bits from oscillations during BP, speculatively flips the associated syndromes, and evaluates short BP passes in parallel. It removes Gaussian elimination; its gross-code circuit-noise comparisons vary physical error rate and parallel implementation, so the runtime benefit is an operating-point-dependent comparison rather than a fixed speedup~\cite{Wang2025BPSF}. Relay-BP instead chains BP legs with different memory strengths to escape nonconvergent states without matrix inversion, reporting improved runtime--accuracy trade-offs and lower logical error rates than the evaluated BP-OSD configurations for bivariate bicycle codes under the circuit-level noise model~\cite{Muller2025}.

Matching supplies another route for selected qLDPC families. Sahay et al.~\cite{Sahay2026Matching} use code symmetries and a cylinder construction to decode bivariate bicycle codes with matching, optionally augmented by belief propagation and over-matching. Their comparisons use code-capacity bit-flip noise; they do not establish a general circuit-noise matching decoder for arbitrary qLDPC codes.

Hardware implementations further expose the parallelism in these algorithms. An FPGA Relay-BP decoder maps the decoding graph onto distinct variable-node and check-node computational units for each vertex in the graph, with a reported 24-nanosecond BP iteration~\cite{Maurer2025}.
At circuit-noise $p=10^{-3}$, the study estimates 480 nanoseconds for a twelve-cycle gross-code window, from about twenty iterations on average; the 40-nanosecond per-cycle figure is amortized.
For a lower-noise configuration, it gives an estimate below 240 nanoseconds for Relay-BP-1 under academic circuit noise at $p<10^{-3}$ and fewer than ten mean iterations.
These source-reported operating points are distinct, and neither is a worst-case or complete readout-to-correction bound. A broader FPGA study compares message passing, filtered OSD, and generalized union-find under fixed cycle budgets. It also introduces a linear-parallel-time systolic Gaussian-elimination primitive and reports that relay message passing provides the strongest speed--accuracy trade-off among the evaluated options~\cite{Maurya2025}. GPU-accelerated BP provides another implementation path: a benchmark of ten fixed iterations on an RTX 4090 takes 43.7 microseconds for a $[[784,24,24]]$ bivariate bicycle code, including host/device transfers. The paper uses 63 microseconds as a comparison target; this does not establish a universal qLDPC deadline or accuracy-matched real-time performance~\cite{Ferraz2025}. These examples show why iteration count and execution cost must be considered together: the algorithm and hardware implementation jointly affect the time per iteration, while the iteration count may be fixed or depend on the syndrome and stopping rule.

Generalized belief propagation~\cite{Old2022} applies region-based inference within a single topological code.
It groups variables and checks into a hierarchical region graph, allowing inference to retain correlations within each region. In the evaluated independent-noise setting, it recovers a subthreshold regime that standard BP does not reach. Here the threshold is the physical error rate below which increasing code size suppresses logical errors; subthreshold performance is the logical error rate achieved below that threshold.
This region hierarchy belongs to the inference algorithm and does not concatenate the code.
Soft readout uses likelihoods derived from analog measurement signals to convey confidence in each measurement outcome, rather than retaining only a binary result. In a distance-three transmon experiment, incorporating this information reduced the extracted logical error rate by up to 6.8 percent for soft minimum-weight matching and also improved the performance of a recurrent neural decoder~\cite{Ali2024Analog}.

\subsubsection{Learned and Optimization-Based Decoders}
\label{sec:eng-ml}
The third group comprises learned or optimization-based decoders that replace matching or belief propagation. AlphaQubit~\cite{Bausch2024} uses a recurrent transformer network and decodes Sycamore surface-code data more accurately than the matching baselines evaluated in that study. Separately, although trained on 25-round examples, it maintained its decoding advantage on simulated memory experiments lasting up to $10^5$ error-correction rounds; this was not a live-QPU feedback demonstration. However, the evaluated implementation did not achieve the processing rate needed to keep pace with a $1\,\mu\mathrm{s}$ syndrome cycle: its reported per-round compute times were approximately one to two orders of magnitude above that target. The Mamba decoder~\cite{Mamba2025} addresses this cost by changing the network structure. It replaces the transformer's self-attention, whose cost grows as the fourth power of the code distance, with a structured state-space block whose cost grows only as the square. On Sycamore data it matches transformer accuracy. A separate simulation with added decoder-dependent depolarizing noise reports a higher distance-three/distance-five crossing point for Mamba than for the Transformer.
Optimization-based methods provide another approach to decoding. Basak et al.~\cite{Basak2026} formulate the search for the most likely individual error consistent with the syndrome as a polynomial optimization problem and solve it through a sequence of increasingly tight sum-of-squares relaxations. Lower relaxation levels reduce computational cost but may return approximate solutions, whereas higher levels approach the exact optimum of this individual-error objective. In their surface- and color-code evaluations, levels two and three approach the exact solution, although their reported runtimes remain higher than those of heuristic decoders.

\subsection{Predecoding Cascades}
\label{sec:predecode}
A predecoding cascade adds a restricted-work stage before a complete decoder.
This stage limits its work through a bounded neighborhood, a small lookup table, or a limited iteration count.
The first stage may correct part of the syndrome, resolve a complete easy case, or update the priors used downstream.
In the evaluated low-noise regimes, local processing removes common short error patterns, leaving fewer detection events for the downstream decoder~\cite{Smith2022,LATTE2025}.
The comparison must include preprocessing overhead, residual traffic, fallback frequency, and the combined logical error rate: neither accuracy nor $L_{\rm e2e}$ improves automatically.
Most cascades reviewed here target the surface code; extensions to qLDPC codes are discussed in Section~\ref{sec:predecoding_qldpc}. Table~\ref{tab:hier1} compares the tasks assigned to the first and downstream stages and the reported performance of these cascades.

\begin{table*}[!t]
\centering
\caption{Single-code predecoding cascades: stage responsibilities and reported results. C1--C5 refer to Section~\ref{sec:bg-constraints}; results retain each study's evaluation conditions.}
\label{tab:hier1}
\scriptsize
\renewcommand{\arraystretch}{1.3}
\begin{tabular}{@{}p{2.3cm}p{3.3cm}p{3.3cm}p{1.5cm}p{4.0cm}@{}}
\toprule
Scheme & First-stage / worker task & Downstream / coordination task & Requirements & Reported performance \\
\midrule
\multicolumn{5}{@{}l}{\emph{Predecoding cascades (Section~\ref{sec:predecode})}}\\
Lazy predecoder~\cite{Delfosse2020} & lazy predecoder & union-find / matching & C4, C5 & all-or-fallback; reduced average traffic, full syndrome on escalation \\
Promatch~\cite{Alavisamani2024} & adaptive prematch & real-time matching & C2, C3 & adaptive preprocessing before bounded real-time matching \\
Clique~\cite{Ravi2022} & cryogenic filter & off-chip matching & C4, C5 & modeled cryogenic filtering of off-chip syndrome traffic \\
Local predecoder~\cite{Smith2022} & cellular-automaton filter & matching & C4 & $p=10^{-3}$ phenomenological; $1000\times$ lower bandwidth, $200\times$ faster matching; $50\%$ more qubits at matched target \\
Pinball~\cite{Pinball2025} & cryo-CMOS primitives & room-temperature decoder & C4, C5 & circuit-noise evaluation; post-layout cryogenic design and modeled interface cost \\
BP partial decoder~\cite{Caune2023Partial} &
High-confidence BP corrections &
matching &
C2, C3 &
Circuit-level surface-code simulations; sparser residual syndrome; $2$--$4\times$ faster matching;
threshold improvement\\
Local-global combination~\cite{Chamberland2022} & 3D CNN local predecoder & matching & C1, C2 & Reduced syndrome density; reduced threshold relative to MWPM alone. \\
Neural predecoder~\cite{Chamberland2026Pre} & 3D CNN & PyMatching (evaluated) & C1, C3 & circuit-noise simulations; measured GPU/CPU compute pipeline \\
QuantiSpect~\cite{Gao2026} & separable 3D CNN & matching & C5 & separable network; parameter and compute-pipeline comparison \\
Scalable neural decoder~\cite{Meinerz2021} & Local neural predecoder & union-find & C2, C3 & local neural cleanup before union-find; simulated accuracy \\
Arqade~\cite{Knapen2026} & generated qLDPC predecoder & shared BP-OSD & C4, C5 & generated primitives; workload-dependent shared-decoder utilization \\
Local syndrome preprocessing~\cite{Fan2025} & local event hints & BP-OSD / Relay-BP & C2 & local priors for downstream BP; code-specific runtime/accuracy study \\
Color-code neural predecoder~\cite{Olle2026ColorPre} & local neural cleanup & Chromobius & C2, C3 & circuit noise, $d=31$, $p=0.003$: model-B $X$-basis LER $347\times$ lower; basis-averaged compute $7.33\times$ faster \\

\bottomrule
\end{tabular}
\end{table*}

\subsubsection{Classical and cryogenic local tiers}
The lazy predecoder~\cite{Delfosse2020} accepts a complete correction only for its restricted class of easy syndromes; otherwise the full syndrome is escalated to union-find or matching.
Its numerical comparisons preserve the correction performance of those downstream decoders in the evaluated setting.
Increasing the accepted-shot fraction lowers average central-decoder demand, but unresolved shots still require full syndrome transmission, and simultaneous fallbacks determine burst bandwidth and provisioning.
This all-or-fallback interface should be distinguished from partial syndrome cleanup, which can change the error distribution seen downstream.

Promatch~\cite{Alavisamani2024} adapts the amount of greedy prematching to the syndrome size that the downstream real-time matcher can handle. This reduces high-weight syndromes to a manageable residual problem while prioritizing local matches. Under the source's uniform circuit-level noise model at $p=10^{-4}$ and $d=13$, Promatch with Astrea reports a logical error rate of $2.6\times10^{-14}$; running the Promatch--Astrea pipeline concurrently with Astrea-G reaches $3.4\times10^{-15}$, matching the evaluated MWPM baseline.

Other local predecoders target the amount of syndrome data sent to the downstream decoder.
A greedy cellular-automaton predecoder~\cite{Smith2022} reports approximately $1000\times$ lower bandwidth and $200\times$ faster matching for a distance-$22$ surface code at $p=10^{-3}$ under phenomenological phase-flip noise with imperfect measurements.
Reaching the same target logical failure probability of $10^{-15}$ requires about $50\%$ more qubits than the optimal-decoder comparison, making the accuracy/resource trade-off explicit.
The Clique predecoder~\cite{Ravi2022} uses cryogenic superconducting logic to suppress local syndrome traffic before off-chip decoding.
Pinball~\cite{Pinball2025} updates the cryogenic line for the rotated surface code.
Running at the four-kelvin stage in a 22-nanometer fully depleted silicon-on-insulator process, it checks predecoding primitives against the syndrome subgraph and corrects common trivial errors before transmission.
It covers space-like, time-like, space-time-like, and hook error classes that the earlier Clique predecoder omits.
Its circuit-noise evaluations compare residual syndrome traffic and logical accuracy, while post-layout analysis estimates cryogenic hardware cost~\cite{Pinball2025}.

Belief propagation can also serve as a partial surface-code decoder rather than a complete qLDPC decoder.
Caune et al.~\cite{Caune2023Partial} apply high-confidence BP corrections before MWPM under circuit-level noise, reducing the downstream matching workload and reporting a threshold increase from $0.94\%$ to $1.02\%$ in their evaluation.
The threshold describes error suppression as code size increases; the application also depends on the logical error rate at its chosen code size and physical error rate. The cascade must therefore be evaluated at that operating point.
Partial corrections can remove useful correlations or commit an incorrect local hypothesis; this risk motivates confidence calibration and a matched-accuracy comparison rather than assuming that fewer defects imply a better decoder.

\subsubsection{Neural local tiers}
Neural networks are a natural fit for the local tier, because the common short error chains form a translation-invariant pattern-recognition problem. The local--global approach~\cite{Chamberland2022} combines a local three-dimensional (3D) convolutional neural network (CNN) predecoder with a downstream global decoder under circuit-level noise. The neural stage corrects local error patterns, followed by syndrome collapse or vertical cleanup to reduce the residual syndrome density before global decoding. These syndrome-processing choices affect both the downstream decoding workload and logical accuracy.
The neural predecoder~\cite{Chamberland2026Pre} is also a fully CNN over the 3D spacetime syndrome volume of the rotated surface code.
It operates blockwise and in parallel, predicts both data-qubit and measurement corrections, and passes a reduced-density residual syndrome to a decoder-agnostic global tier such as PyMatching.
The GPU implementation evaluates both complete syndrome volumes and temporal blocks, with amortized per-round runtimes distinguished from completion of a corrected window.
The study evaluates combined GPU-inference and CPU-matching time as well as logical accuracy. At $d=31$ and circuit-level $p=0.006$, its Model 5 reduces the $X$-basis logical error rate by $4.66\times$ relative to uncorrelated PyMatching. The timing measurements cover the compute pipeline: inference timings disable host--device transfers, and predecoder-to-global-decoder transport is estimated and then omitted~\cite{Chamberland2026Pre}.
QuantiSpect~\cite{Gao2026} targets the resource cost (C5) of this approach.
It replaces dense 3D convolutions with separable spatial, temporal, and grouped spatio-temporal branches. At the same receptive-field size of 13, this factorization reduces the parameter count from 1.80 million to 0.663 million and the per-voxel convolutional multiply-accumulate count by approximately $2.84\times$ relative to the Accurate baseline~\cite{Gao2026}.
An earlier scalable neural decoder~\cite{Meinerz2021} runs a small network as a local predecoder over sliding patches.
Its local predictions remove error patterns before union-find processes the residual syndrome, reducing the work assigned to the global stage.

\subsubsection{Generalizing predecoding to qLDPC codes}\label{sec:predecoding_qldpc}
Most early predecoding work targeted the surface code.
Arqade~\cite{Knapen2026} generalizes it to qLDPC codes.
It automatically constructs a lightweight local tier from a code's detector error model and schedules non-conflicting predecoding primitives in parallel via graph coloring.
Across several code families, its evaluation studies how local acceptance changes shared-decoder utilization; these utilization reductions depend on the input error distribution and are not peak-bandwidth guarantees~\cite{Knapen2026}. FPGA synthesis and resource estimates project support for roughly 1800 bivariate bicycle logical qubits on a single device~\cite{Knapen2026}.

Local syndrome preprocessing~\cite{Fan2025} inserts a lightweight stage that detects likely local error events and steers the priors of a downstream belief-propagation decoder.
The hints guide the initial error probabilities so that BP can converge in fewer iterations, rather than fixing a correction before BP runs. The study evaluates the resulting runtime and logical accuracy against unmodified BP-based decoders on bivariate bicycle codes~\cite{Fan2025}.

Neural predecoding also extends to triangular color codes~\cite{Olle2026ColorPre}.
For superdense syndrome-extraction circuits with circuit-level depolarizing noise at $p=0.003$ and $d=31$, Olle et al. report a $347\times$ lower $X$-basis logical failure rate for model B and a $7.33\times$ basis-averaged compute-pipeline speedup over Chromobius alone.
The comparison includes GPU inference and CPU Chromobius processing, not the full readout-to-correction path.
Improvement relative to this non-optimal downstream decoder does not prove accuracy preservation for arbitrary downstream decoders; at smaller distances the inference overhead can also outweigh the runtime benefit.

\subsection{Windowed, Streaming, and Distributed Decoding}
\label{sec:window}
Windowed and streaming methods organize a continuing syndrome stream; distributed methods divide the work across processing elements.
These choices can be part of one decoder or introduce explicit coordination layers.
They do not require a code hierarchy, and windowing should not be conflated with optional syndrome cleanup. Table~\ref{tab:hier2} compares how these designs divide work and coordinate results, together with their reported performance.

\begin{table*}[!t]
\centering
\caption{Windowed, streaming, and distributed decoding: worker coordination and reported results. C1--C5 refer to Section~\ref{sec:bg-constraints}; timing definitions and evaluation conditions vary across studies.}
\label{tab:hier2}
\renewcommand{\arraystretch}{1.3}
\scriptsize
\begin{tabular}{@{}p{2.3cm}p{3.3cm}p{3.3cm}p{1.5cm}p{4.0cm}@{}}
\toprule
Scheme & First-stage / worker task & Downstream / coordination task & Requirements & Reported performance \\
\midrule
Sliding window~\cite{Huang2023} & $W$-round decode & commit $F$ rounds and advance & C1, C3 & serial windowing; performance depends on $W$, $F$, and noise \\
Fusion Blossom~\cite{Wu2023} & parallel MWPM partitions & fusion of boundary solutions & C1 & partition-and-fuse matching through $d{=}33$ \\
Snowflake~\cite{Chan2024} & local union-find updates & persistent streaming state & C1, C5 & distributed single algorithm; reuses overlap work \\
Modular decoding~\cite{Bombin2023Modular} & local decoding tasks & buffered composition & C1, C3 & fault-distance preservation under stated soundness conditions \\
Parallel-window decoder~\cite{Skoric2022} & concurrent windows & boundary-stitch layer & C1 & ${\sim}$linear to ${>}100$ processes \\
SWIPER~\cite{Viszlai2024} & speculative windows & verify and rollback & C2 & simulated program-runtime reduction of $40\%$ at ${>}90\%$ speculation accuracy \\
Spatially parallel decoder~\cite{Lin2024} & spatial patch windows & merge across patches & C1 & spatial scaling with parallel decoding resources \\
LATTE~\cite{LATTE2025} & FPGA neural local unit & asynchronous CPU block decoder & C1, C2, C4 & FPGA/CPU block-and-seam prototype; lattice-surgery workloads \\
DECONET/\allowbreak HELIOS~\cite{Liyanage2025Net} & per-patch Helios workers & FPGA network $+$ fusion & C1, C2 & $100$ logical qubits; $12.01\,\mu$s/block ($2.40\,\mu$s/round amortized); $0.84\,\mu$s/round inverse throughput; $p=0.001$ phenomenological \\
Windowed ghost protocol~\cite{Turner2025} & ghost-edge passes & global windowed decode & C1, C2 & simulation of correlated transversal-gate decoding and schedule cost \\
\bottomrule
\end{tabular}
\end{table*}
The sliding-window decoder~\cite{Huang2023} processes $W$ qLDPC syndrome rounds, commits corrections for the earliest $F$ rounds, and advances by $F$.
The overlap retains context for boundary decisions while limiting the decoding volume.
A sequential worker is adequate when its service-time distribution satisfies the load and buffering conditions in Section~\ref{sec:background}; slower-than-average windows are compatible with steady-state decoding.
If this worker becomes a bottleneck, parallel execution is an option rather than an assumed prerequisite.
The parallel-window decoder~\cite{Skoric2022} decodes non-overlapping windows concurrently and stitches boundary regions in a second layer, trading additional coordination and feed-forward delay for service capacity.
Fusion Blossom similarly partitions matching work and fuses partial solutions; it is a windowing and parallelization method for MWPM rather than a different decoding primitive~\cite{Wu2023}.

Speculation and spatial partitioning address different sources of decoding delay: dependencies between temporal windows and the size of merged patches.
SWIPER~\cite{Viszlai2024} uses speculative window decoding to predict the dependency between adjacent windows and verify it lazily, reducing simulated fault-tolerant program runtime by about forty percent when speculation accuracy exceeds ninety percent.
Spatially parallel decoding~\cite{Lin2024} partitions merged lattice-surgery patches into overlapping spatial windows; its scaling study varies the number of patches together with parallel decoding resources.

Modular decoding makes the correctness of partitioning explicit.
Theorem~1 of Bomb\'{i}n et al.~\cite{Bombin2023Modular} states that modular composition is sound when every sub-task satisfies local-decoder soundness and the buffering condition.
The latter separates committed corrections from unavailable future data; under these assumptions the composed decoder preserves the protocol's fault distance.
The theorem is an accuracy guarantee under stated conditions, not a measured runtime or a guarantee for arbitrary window boundaries.
Breuckmann et al.~\cite{Breuckmann2017Local} examine local toric-code decoders with finite communication speed, including a hierarchy of local processors.
Their hierarchical local-processor construction explicitly includes finite communication speed, so it is not equivalent to assuming instantaneous communication among a flat collection of workers.
Such processing hierarchies and flat distributed decoders should be distinguished by their actual coordination rules, without implying a concatenated code.

\subsubsection{Streaming decoders that reuse computation across rounds}
Snowflake~\cite{Chan2024} reframes windowing as true streaming for the surface code under circuit-level noise.
It uses a modified union-find decoder whose clusters move toward roots or boundaries as the window rises.
It preserves and reuses the decoder's internal state across cycles, rather than discarding the buffer-region computation that overlapping windows repeat.
Committing one layer per cycle and reusing state avoids repeated buffer work, reducing the processor and power demands of the compared forward-method implementation (C5).
It runs on a nearest-neighbor processor grid; its circuit-noise accuracy and classical resource scaling are distinct from measured end-to-end reaction time.
Snowflake is therefore a streaming single-code design in this subsection, with state reuse and distributed local processing rather than a separate fallback decoder.

\subsubsection{Hybrid and network-integrated designs}
LATTE~\cite{LATTE2025} illustrates a latency-oriented local tier feeding a parallelism-oriented global tier.
Its local tier, the neural local decoding unit, is a heavily quantized 3D CNN on an FPGA that decodes digitized surface-code syndromes locally.
Its confidence gate controls which residual syndromes are forwarded, coupling transmission reduction (C4) to the error distribution and combined-decoder accuracy.
Its global tier, the block decoding system, performs asynchronous block decoding on a multi-socket CPU with a Sparse Blossom decoder.
It partitions the decoding graph across time and space and merges adjacent blocks only along a thin two-dimensional seam, all coordinated by a global dynamic scheduler that returns logical feedback.
LATTE evaluates the combined block/seam pipeline on large workloads, including lattice surgery~\cite{LATTE2025}.
DECONET~\cite{Liyanage2025Net} scales real-time decoding to many interacting logical qubits performing lattice surgery~\cite{Litinski2018}, where merged patches create shared syndromes that extend beyond the scope of an isolated per-patch decoder.
It distributes decoding blocks across a network of FPGAs, using Helios decoders as per-patch local workers, and combines them with intra-node fusion and inter-node parallel windowing as the global tier.
A hybrid tree-grid topology gives $O(\log \ell)$ inter-node latency scaling rather than the $O(\sqrt{\ell})$ scaling of a pure grid, where $\ell$ is the number of logical qubits.
On five \mbox{AMD~Versal\texttrademark{} Prime Series} VMK180 evaluation kits, the DECONET/HELIOS prototype decodes one hundred interconnected distance-five logical qubits with a 12.01-microsecond block $L_{\rm decode}$, measured from the availability of the block's last measurement round at the decoder to the availability of the decoded result, and inverse throughput of 0.84 microseconds per round under 0.1-percent phenomenological noise~\cite{Liyanage2025Net}.
An earlier multi-FPGA prototype establishes the system lineage and isolates the communication problem created by lattice-surgery boundaries~\cite{Liyanage2024MultiFPGA}.
Distributed color-code work extends this question across quantum-processing modules by partitioning color-code patches among modules and evaluating tensor-network and concatenated-matching decoders across noisy intermodule seams~\cite{Chandra2025Color}.
Its modular quantum layout is relevant to distributed decoding even without a near-qubit predecoder and separate global fallback.
The windowed ghost protocol~\cite{Turner2025} addresses fast transversal logic, where transversal gates reduce syndrome rounds but fuse hyperedges across code blocks and break the locality that keeps decoders scalable.
It iteratively decodes the correlation edges introduced by transversal controlled-NOT gates, restoring modularity and supporting a comparison of transversal-gate and lattice-surgery schedules under the study's noise and resource assumptions.
Learning can also coordinate adjacent windows without a separate merging step.
The self-coordinating neural decoder~\cite{Zhang2026Par} trains a recurrent transformer for sliding-window decoding whose adjacent windows agree at their seams without an explicit merging step.
A TPU v6e throughput benchmark reaches less than one microsecond per round amortized across parallel windows through distance twenty-five; this is not the response time of an individual window. Training and accuracy evaluation are demonstrated only through distance seven, including lower logical error rates than the evaluated matching and clustering baselines on Zuchongzhi 3.2 data.
Here seam consistency is learned within one model, illustrating an alternative to an explicit stitching layer.

\subsection{Comparing Single-Code Design Options}
\label{sec:comparison}
Tables~\ref{tab:engines}, \ref{tab:hier1}, and~\ref{tab:hier2} connect complete decoding algorithms to cascade and coordinated-execution designs, identifying where specialization and partitioning can improve the system trade-off.
A complete single decoder is the baseline; a cascade is useful when its reduction in downstream work exceeds preprocessing and communication costs at acceptable accuracy.
The cellular-automaton example shows why accuracy and resource use must be compared together: its large traffic and matching-time reductions require about $50\%$ more qubits at the stated logical-error target~\cite{Smith2022}. Thus a reduction in downstream work is not sufficient by itself to establish a better system design.

Windowing organizes a long syndrome stream into bounded tasks and is distinct from optional syndrome cleanup.
It can use one sequential worker, several equivalent workers, or explicit stitching layers.
Both flat and layered distributed decoders belong in this comparison; communication topology alone does not imply code concatenation.
Fusion Blossom combines parallel partial solutions, Snowflake reuses state across rounds, and the self-coordinating neural decoder learns agreement between adjacent windows~\cite{Wu2023,Chan2024,Zhang2026Par}. They improve processing capacity through parallel execution, avoid repeated overlap computation through state reuse, or remove an explicit seam-processing step through learned coordination.

The relevant timing comparison uses average service capacity, queue recovery, and the end-to-end latency distribution of Section~\ref{sec:background}.
A cascade that always invokes its second stage can still improve this distribution by shortening that stage's work.
Confidence gating additionally changes how often the second stage is used, but its rare-path load and accuracy must be evaluated.
Establishing a complete-system benefit would require comparison at matched accuracy and load. Where a cited study evaluates only part of the processing path, we retain that scope rather than infer an end-to-end benefit.

\section{Hierarchical Codes and Their Decoders}
\label{sec:codes}
Concatenation changes the encoding structure, whereas the two extensions in Section~\ref{sec:engines} organize inference for a fixed code.
We therefore discuss hierarchical code constructions together with their decoding algorithms and interfaces. Table~\ref{tab:families} summarizes how the code setting and decoding interface differ between concatenation, predecoding cascades, and windowed or distributed execution.

\begin{table*}[!t]
\centering
\caption{Decoding organization and information exchange in single-code execution strategies and concatenated codes.}
\label{tab:families}
\footnotesize
\begin{tabular}{@{}p{3.3cm}p{3.2cm}p{4.8cm}p{2.7cm}@{}}
\toprule
Category & Code setting & Decoding organization and interface & Primary considerations \\
\midrule
Predecoding cascades (Section~\ref{sec:predecode}) & fixed code & preprocessing and optional fallback; residual syndrome or prior updates & accuracy, work and traffic reduction \\
Windowed / streaming / distributed (Section~\ref{sec:window}) & fixed code or interacting patches & sequential or parallel workers; persistent state, boundary or seam information where needed & capacity, buffers, communication \\
Concatenated-code decoding (Section~\ref{sec:codes}) & inner and outer encoding layers & layer decoders; hard/soft information, erasures, or outer syndromes & logical reliability, code overhead, interface \\
\bottomrule
\end{tabular}
\end{table*}
The workload of an outer decoder depends on the code and check schedule; ``outer'' does not imply more computation or a more powerful hardware platform.

\subsection{Code Constructions and Decoding Tasks}
\label{sec:code-constructions}
In a concatenated construction, logical qubits encoded by inner code blocks serve as the constituent qubits of an outer code that further suppresses logical errors.
One example concatenates a hypergraph product outer code~\cite{TillichZemor2009} with a distance-five rotated-surface inner code, reaching lower overhead than a bare surface code in the cited code-capacity depolarizing evaluation once the number of logical qubits and the target distance are large enough~\cite{Haruna2025,Pattison2023}.
The small inner block permits enumeration of sector-wise error probabilities, which can then be passed to the outer decoder as described in Section~\ref{sec:soft}.

The same layered structure appears in several other constructions.
Yoked surface codes~\cite{Gidney2023} measure their outer parity checks by lattice surgery, and their decoder exploits soft information to double or quadruple the effective distance.
Concatenated block codes, such as the many-hypercube codes and growing-Hamming constructions, admit blockwise soft-decision decoding, with exact enumeration feasible only for sufficiently small constituent blocks~\cite{Goto2024,Yoshida2024,Poulin2006}.
ConiQ maps many-hypercube concatenation onto neutral-atom arrays through hierarchical logical addressing and a level-wise compiler representation~\cite{Liu2025ConiQ}.
As an adjacent architecture rather than another concatenation, heterogeneous constructions place generalized-bicycle-code memory beside surface-code compute regions and teleport logical data between them~\cite{Viszlai2025GB}.
Concatenated bosonic constructions use a cat inner code that strongly suppresses bit flips, leaving predominantly phase-flip errors for a repetition or classical low-density parity-check outer decoder, or a Gottesman--Kitaev--Preskill (GKP) inner code that emits an analog syndrome that an outer surface code or qLDPC decoder consumes in analog-input mode~\cite{Guillaud2019,Ruiz2024,Noh2021,Berent2023,Borah2025,Raveendran2021,Zhang2022}.
A superconducting experiment realizes the cat-inner variant and reports logical error suppression as the outer repetition distance grows, which shows that the inner layer of a hierarchical code can be a hardware-level bosonic mode rather than a block of qubits~\cite{Putterman2024}.
Parity-code constructions likewise admit both discrete and bosonic concatenations~\cite{Goeller2024Parity}.
Distributed hyperbolic Floquet codes provide a complementary modular, rather than concatenated, substrate based on low-weight measurements across Bell-connected modules~\cite{Sutcliffe2025Floquet}.

\subsection{Decoding Across Inner and Outer Layers}
\label{sec:soft}
Decoding systems can be constructed with separate decoders for inner and outer code layers. The information that flows between the inner and outer decoders may take several forms. In the simplest of cases, hard decisions pass from the inner to outer decoder. However, there are often significant gains in composite logical error rate from identifying erasure errors or passing soft information in the form of a confidence metric between the decoders. Table~\ref{tab:concat} compares the information used by the inner and outer decoders and the reported accuracy of representative constructions.

\begin{table*}[!t]
\centering
\caption{Inner--outer decoding interfaces and reported accuracy in concatenated codes. Noise models and threshold definitions vary across studies; C3 denotes decoding accuracy.}
\label{tab:concat}
\renewcommand{\arraystretch}{1.3}
\scriptsize
\begin{tabular}{@{}p{2.3cm}p{3.3cm}p{3.3cm}p{1.5cm}p{4.0cm}@{}}
\toprule
Scheme & Inner decoding & Outer decoding & Requirements & Reported performance \\
\midrule
Soft-decision decoder~\cite{Poulin2006} & belief propagation & belief propagation & C3 & independent depolarizing threshold $0.1376\to{\geq}0.1885$; exact likelihoods \\
Soft-output decoders~\cite{Meister2024} & soft-output matching / union-find & belief propagation & C3 & phenomenological noise; outer pseudothreshold $0.032\to0.09$ in inner logical error probability \\
Exact inner-lookup decoder~\cite{Haruna2025} & lookup table & BP-OSD & C3 & logical-error suppression in code-capacity simulations \\
Erasure concatenation~\cite{Li2022} & small detection code & MWPM & C3 & converts circuit faults into erasures \\
GKP-inner / qLDPC-outer soft decoder~\cite{Borah2025} & nearest-lattice point & BP-OSD & C3 & round-specific information improves on static priors \\
Learned bidirectional messages~\cite{Wu2026ConcatMP} & neural message passing & neural message passing & C3 & Hamming depolarizing finite-level pseudothreshold $6.5\%\to12.3\%$ \\
qLDPC inner / Algebraic outer~\cite{Wills2026} & Relay-BP & Guruswami-Sudan & C3 & Teraquop performance with less space overhead than two-gross code \\

\bottomrule
\end{tabular}
\end{table*}
In a soft-information interface, an inner-code block decoder passes reliability information, rather than only a hard decision, to the outer-code decoder. This lets the outer decoder distinguish inner-block decisions with different probabilities of failure instead of treating them as equally reliable.
The soft decision decoder~\cite{Poulin2006} showed an early example of the benefit of delaying decisions of what error occurred until information has propagated to the outermost code level. In the construction Poulin studied, the factor graph for the concatenated code forms a tree and information propagates uni-directionally from the inner to outer code. This allows for computation of \textit{exact} marginals with belief propagation and yields globally-optimal maximum-likelihood decoding.
For depolarizing noise on the concatenated five-qubit code, Poulin found that passing soft information raises the reported threshold from $0.1376$ to at least $0.1885$~\cite{Poulin2006}.
The soft-output decoders~\cite{Meister2024} applied this idea to a qLDPC code concatenated over surface-code blocks. The authors estimate inner-block failure probabilities from matching or union-find cluster geometry and supply them to belief propagation at the outer level. This study examined a distance-five inner surface code and a quasi-cyclic lifted product code (QCLP) as the outer qLDPC code. Sampling over 49 rounds of a phenomenological noise model, the outer QCLP pseudothreshold rises from approximately $0.032$ to $0.09$ in the \emph{inner-code logical failure probability}, not in the physical gate error rate~\cite{Meister2024}.
The exact inner-lookup decoder~\cite{Haruna2025} evaluates a concatenated hypergraph-product-over-surface code under code-capacity depolarizing noise, without noisy syndrome extraction.
For the distance-five inner surface code, it constructs a lookup table that enumerates all $2^{25}$ binary error patterns separately for $X$ and $Z$ syndromes, and returns the logical-error probabilities for each. These probabilities are exact within the separately decoded sectors; the method does not jointly optimize the correlated $X/Z$ components of depolarizing noise.
That lookup table supplies per-block priors to an outer BP-OSD decoder, yielding logical-error suppression over the simulated range~\cite{Haruna2025}.
The erasure-concatenation decoder~\cite{Li2022} applies a different multistage construction: small detection codes convert circuit faults into erasures before a three-dimensional cluster-state decoder processes the residual problem.
The decoding stages follow the code concatenation, with erasure flags as their information interface.
The GKP-inner/qLDPC-outer soft decoder~\cite{Borah2025} extends earlier results combining bosonic inner codes with qLDPC outer codes to include circuit-level noise.
The authors study the impact of hard versus soft information and show gains from soft information for a variety of outer qLDPC codes.

Reliability information exchanged across code layers can also be learned by neural networks. Wu et al.~\cite{Wu2026ConcatMP} learn bidirectional soft-message updates on the hierarchical decoding graph.
For $[[15,7,3]]$ Hamming codes and two or three levels of concatenations, they report a finite-level crossing pseudothreshold under depolarizing noise of $6.5\%\to12.3\%$ relative to bidirectional hard-decision decoding. The authors show that their method tolerates location-dependent and correlated noise models in a separate benchmark on the many-hypercube code built from the $[[6,4,2]]$ code.

\subsection{High-Rate Inner Codes and Algebraic Outer Decoding}
\label{sec:algebraic}
Wills et al.~\cite{Wills2026} study a complementary construction with high-rate inner blocks and syndrome-based algebraic outer decoding.
Rather than concatenate over a small, low-rate inner code, it places an algebraic outer code over a high-rate qLDPC inner code, and it supplies the outer code's non-local connectivity through logical operations on the inner blocks rather than through additional hardware.
The case study arranges $[[144,12,12]]$ gross bivariate bicycle blocks in a grid and operates them with the bicycle instruction set and its sacrificial pivot qubit~\cite{Yoder2025,Cross2024}.
This regime highlights the importance of choosing an outer error representation compatible with inner-block failures.
A high-rate inner block encodes many logical qubits at once, so one inner-block logical failure can produce correlated errors across several of them. An outer code must account for that block-level correlation rather than assume independent faults on each encoded qubit.
The construction resolves this by packaging the eleven non-pivot logical qubits of each block into one Galois qudit of dimension $q=2^{11}=2048$, which recasts a correlated multi-qubit failure as a single large-alphabet symbol error and counts any error confined to one block as weight one.
Quantum Reed--Solomon codes with parameters $[[n,n-2(d-1),d]]_q$ then serve as outer codes, and a Galois qudit Shor scheme fault-tolerantly extracts an overcomplete set of checks protected by a time-like Reed--Solomon code.
Their analysis uses minimum-weight outer corrections, exploiting the list-decodable structure of Reed--Solomon codes to correct errors beyond half the outer distance.
Under uniform $10^{-3}$ noise and a budget of 500,000 physical qubits, the authors estimate the required footprint and logical error rate, finding that the concatenated gross system reaches the teraquop regime, roughly $10^{-14}$ to $10^{-13}$ logical error per logical qubit-round, which the bare gross code could not reach at that noise strength, and that it undercuts the space overhead of the $[[288,12,18]]$ two-gross code down to $4.10\times10^{-15}$ per logical qubit-round~\cite{Wills2026}.
These error rates are analytical footprint estimates, not directly observed hardware failure rates. The authors' result rests on retuning the inner Relay-BP decoder, on new compilation strategies, and on decoder post-selection rules, so we read it as a joint code-and-decoder result rather than a property of the code alone~\cite{Wills2026,Muller2025}.

Wills et al.~\cite{Wills2026} make use of a novel method to extract soft information from the inner decoder in order to improve the performance of cat state preparation used to measure outer syndrome checks. 
Forced-gap post-selection~\cite{Wills2026ForcedGap} constructs a confidence metric by forcing one additional decode per logical observable toward the complementary outcome after an initial decode and comparing candidate logical classes. Small gaps trigger rejection in cat state preparation, reducing the logical error rate among accepted outcomes at the cost of additional decoding and retries. A separate manuscript~\cite{Wills2026ForcedGap} evaluates this forced-gap construction for Relay-BP decoding of bivariate bicycle code memories and surgery gadgets under circuit depolarizing noise. Note that as used here, this confidence metric is not supplied as a soft-reliability message to an outer decoder. The results of \cite{Wills2026} pass hard information to an outer Guruswami--Sudan list decoder, leaving the potential benefits of passing soft information to future work.

\subsection{Code and Decoder Co-Design}
\label{sec:code-comparison}
The examples expose two coupled choices: which degrees of freedom an inner block presents to the outer code, and what reliability information the decoding interface retains.
When the inner block is small enough for its error configurations to be summed explicitly, those sums give conditional logical-error probabilities; bosonic blocks can provide analog reliability; high-rate blocks require attention to correlated logical faults~\cite{Poulin2006,Haruna2025,Borah2025,Wills2026}.
Soft outputs can improve accuracy relative to hard decisions, but the benefit and real-time cost depend on the noise model and outer algorithm.
Code and decoder are consequently not independent choices, even when several algorithms are available for the same code.
The algebraic construction also discusses an outer decoder with performance guarantees as a possible response to slow heuristic inner decoding, but does not implement or evaluate such a timing fallback~\cite{Wills2026}.

\section{Real-Time Systems and Implementation}
\label{sec:systems}
Hardware implementation determines how decoding, communication, and control share the available processing and timing budgets. A physical implementation maps the single-code methods of Section~\ref{sec:engines} or the concatenated-code decoders of Section~\ref{sec:codes} onto computing platforms and connects them to readout and control.
One decoder may span several devices, while a cascade may run on one device.
We therefore describe algorithmic stages, placement, and implementation evidence separately in Table~\ref{tab:systems}.

\begin{table*}[!t]
\centering
\caption{Hardware implementations, integration evidence, and reported performance. Timing boundaries differ across studies; decoder and accelerator measurements need not include the full $L_{\rm e2e}$.}
\label{tab:systems}
\renewcommand{\arraystretch}{1.3}
\scriptsize
\begin{tabular}{@{}p{2.5cm}p{2.8cm}p{2.4cm}p{2.4cm}p{5.0cm}@{}}
\toprule
System & Platform placement & Classification & Evidence & Reported result \\
\midrule
Neural predecoder~\cite{Chamberland2026Pre} & GPU local $+$ CPU matching & predecoding cascade & measured compute, simulated syndromes & combined GPU/CPU compute benchmark; circuit-noise simulation; no live QPU \\
Pinball~\cite{Pinball2025} & cryo-CMOS local $+$ room-temperature global & predecoding cascade & post-layout & ${<}0.56$ mW post-layout peak; workload-dependent modeled energy and traffic savings \\
Cryogenic exhaustive search~\cite{Wang2024Cryo} & four-kelvin cryo-CMOS & single decoder & synthesis estimate & $1.13\times10^{-3}$ nJ per logical qubit per correction at $d{=}5$ \\
LATTE~\cite{LATTE2025} & FPGA local $+$ CPU block/seam & cascade $+$ windowed / distributed & FPGA $+$ software prototype & combined block/seam compute and transmission study; lattice-surgery workloads \\
DECONET/\allowbreak HELIOS~\cite{Liyanage2025Net} & five networked FPGAs & windowed / streaming / distributed & measured FPGA prototype & $12.01\,\mu$s/block ($2.40\,\mu$s/round amortized); $0.84\,\mu$s/round inverse throughput; $p=0.001$ phenomenological \\
ZeroG~\cite{ZeroG2026} & CPU or FPGA strong tier after predecode & single strong tier; predecoder-aware & CPU estimate; measured FPGA kernel only & $0.93\,\mu$s FPGA typical and $1.45\,\mu$s 99th percentile at $d{=}15$ \\
THQLink~\cite{THQLink2026} & FPGA endpoint $+$ CPU cluster & windowed / streaming / distributed & measured classical loop & $2.944\,\mu$s raw network round trip; sustains $1\,\mu$s rounds through $d{=}19$ \\
Willow~\cite{GoogleWillow2025} & QPU/control $+$ CPU decoder & windowed / streaming / distributed & live syndrome stream & $63\pm17\,\mu$s decoder-only terminal latency at $d{=}5$; $1.1\,\mu$s cycles \\
Deltaflow 2~\cite{Deltaflow2026} & QPU/control emulator $+$ FPGA system & streaming / windowed & hardware replay of QPU data; vendor report & $16.32\,\mu$s mean terminal response ($8.28\,\mu$s SD), last data sent to returned result \\
Quantinuum Helios~\cite{NVIDIAHelios2025,NVIDIAHeliosNews2025} & trapped-ion QPU $+$ GH200 GPU & single BP+OSD decoder & live-QPU feed-forward, vendor report & $67\,\mu$s median decode; $2$ ms stated target; full C2 boundary unspecified \\
Analog-readout decoding~\cite{Ali2024Analog} & superconducting QPU $+$ software decoder & single decoder, soft input & measured QPU data & up to $6.8\%$ lower extracted logical error rate \\
Collision-clustering feedback~\cite{Caune2026} & QPU/control $+$ FPGA decoder & single decoder & live-QPU feedback & $9.6\,\mu$s for nine rounds, including $3.1\,\mu$s control/communication \\
FPGA neural feedback~\cite{Yang2026FPGA} & QPU/control $+$ FPGA neural decoder & single decoder & live-QPU feedback & $124$ ns decode; $550$ ns deterministic readout-completion-to-feedback path at $d{=}3$ \\
Greedy ASIC decoder~\cite{Kadomoto2025} & fabricated 22-nm CMOS & single decoder & measured silicon & $0.0235$ mm$^2$; $0.16$ mW measured at $10$ MHz \\
\bottomrule
\end{tabular}
\end{table*}

\subsection{End-to-End Timing and Scheduling}
\label{sec:systems-timing}
Equation~\ref{eq:e2e-latency} defines the readout-to-correction path used here.
Each implementation should identify which parts it actually measures: a kernel benchmark excludes acquisition, link traffic, waiting, and control unless explicitly included.
Dividing a window runtime by its round count gives an amortized quantity, not the elapsed time for the window's corrected decision.
For overlapping windows, effective throughput depends on the newly committed rounds rather than the full window width, as discussed under C1 in Section~\ref{sec:background}.

Latency distributions connect implementation behavior to scheduling.
A low mean processing time supports throughput, but expensive syndromes, batching, or network contention can temporarily delay incoming work and cause queues to grow.
The relevant question is whether that queue recovers within the available buffers and logical-operation schedule, not whether every window meets one fixed service bound.
At a feed-forward point, the remaining readout-to-correction work may extend the logical operation; a hard deadline is an additional protocol assumption.

DART-Q~\cite{DARTQ2026} studies one deadline-driven scheduling formulation using earliest-deadline-first tasks, admission control, and decoder-state placement in fast memory.
A study of large-depth circuits~\cite{Khalid2025} quantifies how combined decoder and control reaction time changes architectural overhead and runtime for workloads with many non-Clifford gates.
The system-stack analysis~\cite{RTStack2026} distinguishes Pauli-frame tracking from more time-sensitive feed-forward paths.
These perspectives motivate reporting means, high percentiles, queue growth and recovery, and schedule-dependent stall costs together, rather than treating a tail event as automatic failure.
For a broader discussion of real-time decoding requirements and implementation challenges, see Battistel et al.~\cite{Battistel2023Review}.

In cascades with fallback decoding, the distribution of readout-to-correction latency ($L_{\mathrm{e2e}}$) reflects both cases resolved by the initial stage and those requiring the fallback decoder.
A predecoder that always invokes a downstream decoder can still reduce end-to-end latency by shrinking its work.
Confidence gating also reduces how often that stage runs, but difficult bursts may concentrate escalations and cause contention~\cite{ConfGated2026,Chamberland2026Pre}.
A controlled evaluation should compare the cascade with a single decoder at matched logical accuracy and load; this is an evaluation criterion, not a condition met by every reported result in this paper.

\subsection{Platform-Specific Realizations}
\label{sec:systems-platforms}

\subsubsection{CPU and GPU}
CPUs offer mature memory systems, flexible graph processing, and rapid adaptation to a changing code or noise model.
Sparse Blossom and Fusion Blossom demonstrate that optimized and parallel CPU matching can sustain large syndrome streams~\cite{Higgott2023,Wu2023}.
This flexibility comes with operating-system, memory, and network variability, so an amortized time per round does not imply the same terminal response time.
GPUs expose parallelism for belief propagation, learned filters, and concurrent decoding windows.
The GPU belief-propagation benchmark~\cite{Ferraz2025} averages 43.7 microseconds for ten fixed iterations on a $[[784,24,24]]$ qLDPC code, including host/device transfers, while the neural predecoder~\cite{Chamberland2026Pre} uses GPU inference to reduce the residual problem passed to CPU matching.
The latter is a genuine heterogeneous hierarchy, but its blockwise measurements characterize compute throughput rather than a live quantum-processor feedback loop.

CUDA-Q QEC 0.6 also supplies a real-time integration path through NVQLink and \texttt{cudaq-realtime}, with RelayBP and a surface-code CNN predecoder followed by CPU PyMatching~\cite{CUDAQ06}.
The CNN is deployed through TensorRT in FP8 or FP16; incoming syndromes are batched into a pool of CUDA graphs and unresolved cases proceed to matching.
This is a capability of the software and runtime interface; throughput measurements of batched decoding should still be distinguished from complete live-QPU readout-to-correction latency.

The IonQ MegaQuOp study~\cite{Ye2026MegaQuOp} benchmarks dual sliding-window beam-search decoders and on-the-fly detector-error-model generation on twelve cores of one Apple M4 Max CPU. Its simulated trapped-ion workloads include up to 408 logical qubits and over one million $T$ gates in the largest $T$-count workload. With assumed 1--5 ms syndrome cycles, decoding-induced schedule extension remains below $0.3\%$ at $p_{\rm CNOT}=10^{-4}$ and below $12\%$ at $5\times10^{-4}$. This demonstrates workload-level CPU feasibility under those timing and noise assumptions, including the cost of slow windows, rather than a measured readout-to-correction loop with quantum hardware.

\subsubsection{FPGA}
FPGAs trade software flexibility for pipelines with controllable timing, direct control-system interfaces, and fine-grained parallelism.
They host exact matching in Micro Blossom, clustering in Helios and Collision Clustering, Relay-BP for qLDPC codes, and neural inference in LATTE~\cite{Wu2025,Liyanage2024,Liyanage2023QCE,Barber2023,Maurer2025,LATTE2025}.
The resource model differs across these designs.
Graph-parallel decoders replicate processing elements with code size, neural decoders consume multipliers and on-chip memory, and lookup-table designs exchange compute for rapidly growing storage.
Multi-context systolic arrays instead time-multiplex union-find state to reduce the number of processing elements~\cite{Heer2023}.
Branchini et al. review these FPGA mapping choices and their latency, parallelism, memory, and resource trade-offs~\cite{Branchini2023}.
Changes to supported weights or noise parameters can be handled without changing the algorithmic architecture; changing the graph, precision, or algorithm is a separate implementation question. Noise-model updates therefore do not intrinsically require a CPU or GPU. The platform can implement a complete reaction-path decoder or a stage of a larger design. At larger distances, memory, logic, and communication costs determine whether temporal reuse or multiple devices are useful.

\subsubsection{ASIC and Cryogenic Logic}
ASICs can provide low projected $L_{\rm decode}$ and energy for a selected code and algorithm~\cite{Barber2023,Kadomoto2025}. Updating weights within a supported model may remain possible; changing the algorithm or supported graph is a different and potentially more costly modification.
The collision-clustering design reports post-layout results of 240 nanoseconds per round, amortized over the decoded rounds, and approximately 8 milliwatts at distance twenty-three~\cite{Barber2023}.
These figures are implementation estimates rather than measurements from fabricated decoder silicon.
The greedy ASIC decoder~\cite{Kadomoto2025} provides a complementary evidence point from a fabricated 22-nanometer chip.
The distance-fifteen core occupies 0.0235 square millimeters, and the authors functionally verified it at 10 megahertz, where a 1275-cycle decode takes 127.5 microseconds and consumes 0.16 milliwatts.
Its 12.75-microsecond $L_{\rm decode}$ and 1.6-milliwatt power at 100 megahertz are post-layout projections.

Cryogenic preprocessing is one way to reduce syndrome traffic before it leaves a low-temperature stage. Link engineering and readout-side aggregation are alternative system choices; a bandwidth limit does not itself require cryogenic decoding.
Clique uses superconducting logic for common local errors, whereas Pinball maps a broader circuit-level primitive set into 22-nanometer fully depleted silicon-on-insulator cryo-CMOS~\cite{Ravi2022,Pinball2025}.
Pinball reports a post-layout peak power below 0.56 milliwatts; its energy and transmission savings depend on the modeled syndrome distribution and interface overheads~\cite{Pinball2025}.
The cryogenic exhaustive-search decoder estimates $1.13\times10^{-3}$ nanojoules per logical qubit per correction for distance five at four kelvin~\cite{Wang2024Cryo}.
The reported energy is a synthesis estimate for the proposed cryogenic circuit; fabricated-silicon power and timing measurements are not reported.
Arqade generalizes this placement to qLDPC codes and projects roughly 50,000--500,000 bivariate bicycle logical qubits under a 1.5-watt cryogenic budget~\cite{Knapen2026}.
These cryogenic results are synthesis or post-layout studies, not fabricated predecoders connected simultaneously to a quantum processor and a room-temperature global decoder.

\subsection{Multi-Stage and Networked Systems}
\label{sec:systems-heterogeneous}
The following systems explore how explicit stages, parallel workers, and networked coprocessors realize specialization and coordinated execution. Their measurements expose the compute, communication, and scheduling trade-offs of these designs.
LATTE places a quantized neural local tier on an FPGA and asynchronous Sparse Blossom blocks on a multi-socket CPU; the local tier reduces both defect density and transmission, while the CPU tier supplies flexible spatial and temporal merging~\cite{LATTE2025}.
DECONET instead distributes per-patch Helios workers and global fusion across five AMD Versal Prime Series VMK180 evaluation kits.
The DECONET/HELIOS prototype evolves from an earlier multi-FPGA lattice-surgery system and decodes one hundred interacting distance-five logical qubits with a 12.01-microsecond block $L_{\rm decode}$ and inverse throughput of 0.84 microseconds per round~\cite{Liyanage2024MultiFPGA,Liyanage2025Net}.

ZeroG~\cite{ZeroG2026} co-designs the strong tier for the sparse residual syndromes produced by a predecoder.
For ZeroG alone, a multiplexing estimate supports up to 640 logical qubits on 128 CPU cores under a one-microsecond budget; the source notes that adding predecoder latency reduces this capacity~\cite{ZeroG2026}.
The measured configuration evaluates the predecoder-aware strong tier rather than a complete predecoding cascade.
The design uses a high-bandwidth memory (HBM) adaptive system-on-chip (SoC).
On an \mbox{AMD~Alveo\texttrademark{} V80 accelerator card}, powered by an AMD Versal HBM adaptive SoC, the measured kernel averages 0.93 microseconds for typical distance-fifteen synthetic residuals, with a 1.45-microsecond 99th percentile; a density-oriented configuration supports 32 decoder instances.
The CPU capacity estimate and FPGA kernel measurements answer different questions: how many logical qubits can share the processor, and how long one residual decoding task takes. They therefore cannot be read as a direct platform speedup comparison.

THQLink~\cite{THQLink2026} extends the heterogeneous pattern across a high-performance-computing interconnect.
An FPGA emulates the quantum-control endpoint, and a CPU cluster runs matching over parallel data windows followed by seam windows.
The measured raw network round trip averages 2.944 microseconds, each additional switch hop adds about 130 nanoseconds, and the windowed pipeline sustains one-microsecond syndrome rounds through distance nineteen~\cite{THQLink2026}.
This is a closed classical hardware loop rather than an experiment involving a live quantum processing unit (QPU), but it exposes transport and switching costs that accelerator-only measurements omit.

Backline~\cite{Backline2026} extends this networked pattern by providing a heterogeneous compilation
and runtime framework, built within PennyLane quantum software platform~\cite{PennyLane2018} and
Catalyst compiler~\cite{Catalyst2024}, that enables the design and compilation of quantum-classical
workloads for high-performance and low-latency devices directly from a Python interface through LLVM
and MLIR.
Operating primarily within what the authors define as the synchronous coprocessing tier (1--100 microseconds)
for soft real-time tasks like backup decoding, the framework characterizes the transport costs of mapping
coprocessor functions to commodity targets. In its experimental setup, an AMD Versal\texttrademark{}
VPK120 FPGA controller issues 16-byte payloads via its hardware-handshake engine over RoCE v2.
Baseline steady-state round-trip latencies measured 2.305 microseconds to an AMD Ryzen\texttrademark{}
Threadripper PRO CPU and 4.5 microseconds to an AMD Instinct\texttrademark{} MI210 GPU. For
simple tasks like Steane lookup decoders, execution time is virtually free compared to communication
overhead, yielding total median round-trip times (2.305 microseconds for CPU, 4.405 microseconds for
GPU) that closely track the baseline network echo. Against a deterministic reaction budget, the CPU
path proved substantially more predictable with a standard deviation of just 21.4 nanoseconds, compared
with a 159-nanosecond spread for the GPU. This work shows that commodity CPUs present a strong
path for highly predictable outer-tier decoding workloads.

The baseline values above are echo medians, and the quoted standard deviations describe that baseline workload. These controller-to-coprocessor measurements do not include a complete QPU readout-to-correction path or establish a worst-case timing guarantee.

\subsection{Integration with Quantum Hardware}
\label{sec:systems-qpu}
Live QPU experiments establish whether readout, decoding, and control can operate together.
The analog-readout decoding experiment uses data from a superconducting logical qubit to improve both matching and neural decoding~\cite{Ali2024Analog}.
The experiment validates a soft-input decoder on measured data, although it does not close a conditional feedback loop.
Google's distance-five Willow experiment sustains 1.1-microsecond syndrome cycles for up to one million rounds with a mean decoder-only terminal latency of $63\pm17$ microseconds after receipt of the final syndrome~\cite{GoogleWillow2025}.
Its real-time decoder assigns ten-cycle spacetime blocks to separate threads and fuses boundary detections downstream, so it is a window-and-fusion hierarchy.
The system demonstrates backlog-free streaming, but transmission adds further latency, and the experiment did not implement physical feed-forward.

Riverlane's Deltaflow 2 replays the recorded distance-five Willow measurements through a QPU/control emulator and its QECi interface, with FPGA decoding and return of corrected results~\cite{Deltaflow2026}.
It reports a 16.32-microsecond mean terminal response and 8.28-microsecond standard deviation across ten shots.
The boundary is the last data sent by the emulator to the returned corrected result; it includes interface and processing overheads but excludes a live quantum readout and physical actuation.
This result demonstrates hardware processing of experimental data, not a live-QPU feedback loop.
The source explicitly cautions that comparison with Willow's decoder-only 63-microsecond result is not strictly like-for-like.

Quantinuum's Helios trapped-ion processor provides a complementary GPU example, distinct from the similarly named FPGA Helios decoder above.
NVIDIA reports Bring's code, encoding eight logical qubits into thirty physical qubits, decoded by BP+OSD on a GH200 Grace Hopper host, with a 67-microsecond median decoding time and real-time feed-forward corrections~\cite{NVIDIAHelios2025}.
A companion report states a two-millisecond Helios requirement~\cite{NVIDIAHeliosNews2025}.
We classify this as vendor-reported live-QPU feed-forward; the technical account identifies the 67-microsecond statistic as decoding time and does not specify a complete readout-to-actuation timing boundary.
The demonstration establishes feasibility for its trapped-ion workload, without implying that a near-qubit preprocessing tier is universally necessary for faster syndrome streams.

The collision-clustering feedback system~\cite{Caune2026} integrates an FPGA decoder into the control system of the Rigetti Ankaa-2 processor.
For nine rounds, the measured 9.6-microsecond response comprises 6.5 microseconds of decoding and 3.1 microseconds of communication and control.
The mean decode time remains below one microsecond per round through twenty-five rounds, and the experiment performs conditional physical feedback.
The FPGA neural feedback system~\cite{Yang2026FPGA} reports a tighter loop for a distance-three surface code: a quantized neural decoder on an FPGA takes 124 nanoseconds, while the complete path from readout completion to feedback pulse takes 550 nanoseconds within a 1.25-microsecond cycle.
It also demonstrates mid-circuit feedback for a non-Clifford logical rotation using a non-fault-tolerant rotation protocol; this is not a demonstration of universal fault-tolerant logical computation.
Both systems show that a single decoder can participate in a measured decoding-and-feedback loop. Physical Pauli feedback and use of a corrected logical measurement in fault-tolerant adaptive computation should still be reported separately.

\subsection{Adjacent High-Performance Computing Infrastructure}
Evaluating a decoding system also requires a quantum-operation schedule, a communication model, and a way to generate representative workloads. The following tools support those inputs rather than directly measure decoder performance.
The trapped-ion surface code compiler (TISCC) lowers repeated error-correction rounds and lattice-surgery operations to ion-motion schedules and reports their space-time cost~\cite{LeBlond2023TISCC}.
The quantum message passing interface (QMPI) and DQTetris characterize communication in distributed quantum architectures, while Qonductor addresses heterogeneous quantum-classical resource orchestration~\cite{Haner2021QMPI,Xu2025DQTetris,Giortamis2025Qonductor}.
UniQ and Atlas provide heterogeneous and hierarchically partitioned simulation infrastructure that can support large encoded-circuit studies~\cite{Zhang2022UniQ,Xu2024Atlas}.
Work on quantum Fourier transformation kernels further exposes the lattice-surgery workloads that a scalable decoder must follow~\cite{Jin2024QFT}.
For decoder evaluation, these tools can help specify when syndrome data arrives, which operations require corrected results, and what communication the target architecture permits. Their performance results should not be interpreted as decoder benchmarks.

\subsection{Cross-Platform Lessons}
\label{sec:systems-lessons}
Hardware choice follows the workload and its timing budget.
FPGA and ASIC pipelines can reduce jitter, but data-dependent iteration counts, queues, and links can still vary; fixed-cycle kernel timing does not guarantee a deterministic full system.
A fixed-depth FPGA or ASIC kernel can have a bounded execution time for a specified input size, while iterative hardware needs an explicit iteration cap and associated failure policy.
An end-to-end bound additionally requires bounded queues, arbitration, transport, and control response.
CPU core pinning and real-time scheduling, or GPU persistent kernels and preallocated graph dispatch, can reduce jitter; their use alone does not prove a worst-case guarantee.
Among the cited demonstrations, the FPGA neural feedback experiment reports a deterministic 550-nanosecond path from readout completion to feedback for its particular configuration~\cite{Yang2026FPGA}; this is not a guarantee for every FPGA decoder.
CPU and GPU implementations offer flexible scheduling and large memories, and may suit both complete decoders and pipeline stages.
Their suitability depends on the measured $L_{\rm e2e}$ distribution, syndrome cadence, window advance, and feed-forward schedule, which vary across modalities.
Microsecond superconducting rounds should not be treated as a universal reaction-time deadline or a universal ban on GPU decoding.
Table~\ref{tab:modalities} connects these choices to physical-platform requirements: syndrome cadence determines incoming processing demand, while readout overhead and dependent-operation schedules constrain the available response time.
The reviewed evidence is uneven across modalities and does not establish a preferred classical platform for each one. Energy and complete readout-to-correction measurements are also unavailable for many implementations, limiting comparisons on those metrics.

Communication is part of the readout-to-correction path.
The 3.1-microsecond control/communication contribution in the Rigetti--Riverlane loop and the 2.944-microsecond raw THQLink round trip illustrate costs omitted from kernel measurements~\cite{Caune2026,THQLink2026}.
Compare these costs to the window schedule and full feedback budget, not only to the duration of one syndrome round.

Implementation evidence must also be explicit.
Simulation tests accuracy under a noise model, synthesis and post-layout analysis estimate feasibility and cost, and hardware measurements establish behavior on a particular device.
Live syndrome ingestion, offline replay of QPU data, physical feedback, and fault-tolerant logical feed-forward establish different levels of integration.
Neither a small average runtime nor the choice of platform alone establishes worst-case execution time.

\section{Challenges and Open Problems}
\label{sec:challenges}
Several obstacles separate the reviewed decoders from a deployable real-time FTQC system.

\emph{Latency distributions, buffers, and feed-forward.}
Average service capacity, queue recovery after a burst, and end-to-end response at dependent operations should be evaluated together.
The open problem is how to provision decoding and transport so that long-tail work remains compatible with finite buffers and the application's error and runtime budgets.
Single decoders, parallel workers, bounded-iteration algorithms, and gated cascades are alternative approaches; a local tier that resolves every easy case is not a prerequisite for improving $L_{\rm e2e}$~\cite{Skoric2022,Muller2025,ConfGated2026,RTStack2026}.

\emph{Real-time soft information in hardware.}
Passing soft information between tiers improves accuracy in the cited software studies, and measured analog readout improves small-code decoding~\cite{Ali2024Analog}.
The implementation question is how generating, transporting, and consuming these probabilities affects the full $L_{\rm e2e}$ distribution and memory cost at the target block size~\cite{Meister2024,Haruna2025,Borah2025}. A software accuracy result alone does not establish that timing feasibility.

\emph{Scaling to many logical qubits and lattice surgery.}
Real-time decoding of an isolated memory patch is comparatively mature.
Decoding hundreds of interacting logical qubits under lattice surgery, where merges create shared syndromes and dynamic decoding graphs, has fewer measured demonstrations. Beyond memory, the IonQ MegaQuOp study evaluates a complete software decoding and logical-control stack for compiled trapped-ion computation, including adaptive logical measurements and magic-state factories~\cite{Ye2026MegaQuOp}. Its walking-cat architecture preserves the bulk Tanner graph, so this result is complementary to the dynamic-graph lattice-surgery problem; it is a classical benchmark driven by sampled errors, not a live quantum-computer demonstration.
The distributed and network-integrated designs that address it face communication-latency and synchronization limits~\cite{Skoric2022,Lin2024,Liyanage2024MultiFPGA,Liyanage2025Net,Chandra2025Color,Turner2025}.

\emph{Real-time decoding of qLDPC codes.}
BP-OSD provides a common accuracy baseline, but Gaussian elimination can dominate its processing time as the decoding problem grows.
The next systems question is how the recent alternatives perform over sustained syndrome streams, including transport, queueing, and logical feed-forward, either as complete decoders or as stages within a pipeline~\cite{Muller2025,Maurer2025,Ferraz2025,Delfosse2025}.
The concatenated algebraic memory makes the resulting tension concrete: its reported error rates depend on a Relay-BP configuration that searches one hundred candidate solutions, and the authors note that this setting falls outside the scope of the published FPGA real-time decoders while remaining plausible for an ASIC~\cite{Wills2026,Maurer2025,Maurya2025}.
The authors also note that their memory analysis does not require decoding to keep pace strictly with syndrome extraction because software can correct past mistakes, whereas a computation with feed-forward operations cannot generally tolerate the same backlog.
The allowable lag and its consequences must therefore be specified for the application; feed-forward can tolerate some waiting, but with a schedule-dependent cost.

\emph{Benchmarking heterogeneity.}
As Tables~\ref{tab:hier1}, \ref{tab:hier2}, and~\ref{tab:systems} make clear, decoders report thresholds, latencies, and throughput under different noise models, codes, hardware, and timing boundaries, which prevents direct cross-study comparison.
A shared benchmark suite should report the syndrome source, accelerator and host, batch size, transport path, $L_{\rm e2e}$ distribution, sustained throughput, power, area, energy per defined decoding task, and whether the experiment used offline data, live syndromes, physical Pauli feedback, or corrected logical outcomes to control subsequent logical operations.
A common circuit-level baseline should support controlled comparisons. For superconducting-inspired comparisons, SI1000 assigns different error strengths to gates, measurement, reset, and idling, providing a more specific baseline than uniform depolarizing noise~\cite{Gidney2021Honeycomb}; it is not a universal noise model for every modality. Modality-appropriate extensions can then test leakage, correlated readout, bursts, and drift where relevant, without requiring every experiment to model all of them~\cite{Brown2020Leakage,Vallero2024,Robertson2023,Patel2020NISQ}.
Shared traces and evidence labels would let the community compare algorithms without treating simulation, post-layout estimates, and live-QPU measurements as equivalent~\cite{RTStack2026}.

\emph{Demonstrating logical feed-forward at scale.}
Beyond memory experiments and physical-feedback proxies, systems should demonstrate corrected logical outcomes controlling relevant subsequent logical operations.
For a multi-stage decoder this includes both common and escalated paths; a single-decoder implementation remains an equally relevant baseline.
The experiment should relate $L_{\rm e2e}$ excursions to stalls and logical errors rather than assume that every excursion is fatal.

\section{Conclusion}
\label{sec:conclusion}
Scalable fault-tolerant quantum computing requires careful codesign of the error-correcting code, classical decoder, and the quantum control stack. Real-time QEC decoding is a nontrivial task that must balance logical accuracy, sustained throughput, end-to-end latency, communication, and hardware cost. For a syndrome stream, average processing margin, bounded tails in end-to-end readout-to-correction latencies, and buffering together determine whether decoding can sustain the workload.
When a logical operation needs a corrected result, the relevant delay includes readout, transport, waiting, decoding, and control.

We organized the decoding literature around two code settings, followed by a discussion of hardware implementation.
For a fixed code, complete single decoders form the baseline for exploring how specialized stages and windowed, streaming, or distributed execution can reduce downstream work, organize concurrency, and improve resource use.
For hierarchical codes, inner and outer decoding must be considered together with the encoding layers and their hard, soft, erasure, or syndrome interfaces.
Together, these settings identify two opportunities for hierarchical decoding: organizing the execution of a decoding task and using information across the layers of an encoded system.

The hardware discussion relates both settings to qubit-modality requirements and CPU, GPU, FPGA, ASIC, and cryogenic implementations, retaining the timing boundaries and evaluation conditions of the reported results.

By connecting decoding mechanisms to their reported accuracy, timing, communication, and resource costs, this paper helps readers identify relevant design options and the conditions under which their benefits have been demonstrated. It also identifies gaps between component measurements and complete-system evidence that future evaluations need to address.

\section*{Acknowledgment}
We thank the colleagues and collaborators who suggested decoders for inclusion and who sharpened earlier drafts of the taxonomy and the comparison tables.

AMD, the AMD Arrow logo, Alveo, Instinct, Ryzen, Threadripper, Versal and combinations thereof are trademarks of Advanced Micro Devices, Inc.
Other product names used in this publication are for identification purposes only and may be trademarks of their respective companies.

\bibliographystyle{IEEEtran}
\bibliography{references}

\end{document}